\documentclass[preprint]{aastex61}
\usepackage{amsmath}
\usepackage{natbib}
\usepackage{graphicx}
\usepackage{multirow}
\usepackage{epstopdf}
\usepackage{mathrsfs}
\usepackage{color}
\begin{document}

\title{Searching for the Transit of the Earth-mass exoplanet Proxima~Centauri~b in Antarctica: Preliminary Result}
\author{Hui-Gen Liu}
\affiliation{School of Astronomy and Space Science, Nanjing University, Nanjing, Jiangsu, 210093, China}
\affiliation{Key Laboratory of Modern Astronomy and Astrophysics (Nanjing University), Ministry of Education, Nanjing, Jiangsu, 210093, China}

\author{Peng Jiang}
\affiliation{Polar Research Institute of China, Jinqiao Rd. 451, Shanghai, 200136, China}
\affiliation{Key Laboratory for Research in Galaxies and Cosmology, The University of Science and Technology of China, Chinese Academy of Sciences,
Hefei, Anhui, 230026, China}

\author{Xingxing Huang}
\affiliation{Department of Astronomy, The University of Science and Technology of China, Hefei, Anhui, 230026, China}
\affiliation{Key Laboratory for Research in Galaxies and Cosmology, The University of Science and Technology of China, Chinese Academy of Sciences,
Hefei, Anhui, 230026, China}

\author{Zhou-Yi Yu}
\affiliation{School of Astronomy and Space Science, Nanjing University, Nanjing, Jiangsu, 210093, China}
\affiliation{Key Laboratory of Modern Astronomy and Astrophysics (Nanjing University), Ministry of Education, Nanjing, Jiangsu, 210093, China}

\author{Ming Yang}
\affiliation{School of Astronomy and Space Science, Nanjing University, Nanjing, Jiangsu, 210093, China}
\affiliation{Key Laboratory of Modern Astronomy and Astrophysics (Nanjing University), Ministry of Education, Nanjing, Jiangsu, 210093, China}

\author{Minghao Jia}
\affiliation{Department of Modern Physics, The University of Science and Technology of China, Hefei, Anhui, 230026, China}
\affiliation{State Key Laboratory of Technologies of Particle Detection and Electronics, Department of Modern Physics, University of Science and
Technology of China, Hefei, Anhui, 230026, China}

\author{Supachai Awiphan}
\affiliation{National Astronomical Research Institute of Thailand, 191 Huay Kaew Road, Amphoe Mueang, Chiang Mai, Thailand 50200}

\author{Xiang Pan}
\affiliation{Department of Astronomy, The University of Science and Technology of China, Hefei, Anhui, 230026, China}
\affiliation{Key Laboratory for Research in Galaxies and Cosmology, The University of Science and Technology of China, Chinese Academy of Sciences,
Hefei, Anhui, 230026, China}
\affiliation{Polar Research Institute of China, Jinqiao Rd. 451, Shanghai, 200136, China}

\author{Bo Liu}
\affiliation{Department of Astronomy, The University of Science and Technology of China, Hefei, Anhui, 230026, China}
\affiliation{Key Laboratory for Research in Galaxies and Cosmology, The University of Science and Technology of China, Chinese Academy of Sciences,
Hefei, Anhui, 230026, China}
\affiliation{Polar Research Institute of China, Jinqiao Rd. 451, Shanghai, 200136, China}

\author{Hongfei Zhang}
\affiliation{Department of Modern Physics, The University of Science and Technology of China, Hefei, Anhui, 230026, China}
\affiliation{State Key Laboratory of Technologies of Particle Detection and Electronics, Department of Modern Physics, University of Science and
Technology of China, Hefei, Anhui, 230026, China}

\author{Jian Wang}
\affiliation{Department of Modern Physics, The University of Science and Technology of China, Hefei, Anhui, 230026, China}
\affiliation{State Key Laboratory of Technologies of Particle Detection and Electronics, Department of Modern Physics, University of Science and
Technology of China, Hefei, Anhui, 230026, China}

\author{Zhengyang Li}
\affiliation{National Astronomical Observatories/Nanjing Institute of Astronomical Optics and Technology, Chinese Academy of Sciences, Nanjing,
Jiangsu, 210042, China}

\author{Fujia Du}
\affiliation{National Astronomical Observatories/Nanjing Institute of Astronomical Optics and Technology, Chinese Academy of Sciences, Nanjing,
Jiangsu, 210042, China}

\author{Xiaoyan Li}
\affiliation{National Astronomical Observatories/Nanjing Institute of Astronomical Optics and Technology, Chinese Academy of Sciences, Nanjing,
Jiangsu, 210042, China}

\author{Haiping Lu}
\affiliation{National Astronomical Observatories/Nanjing Institute of Astronomical Optics and Technology, Chinese Academy of Sciences, Nanjing,
Jiangsu, 210042, China}

\author{Zhiyong Zhang}
\affiliation{National Astronomical Observatories/Nanjing Institute of Astronomical Optics and Technology, Chinese Academy of Sciences, Nanjing,
Jiangsu, 210042, China}

\author{Qi-Guo Tian}
\affiliation{Polar Research Institute of China, Jinqiao Rd. 451, Shanghai, 200136, China}
\affiliation{Key Laboratory for Research in Galaxies and Cosmology, The University of Science and Technology of China, Chinese Academy of Sciences,
Hefei, Anhui, 230026, China}

\author{Bin Li}
\affiliation{Polar Research Institute of China, Jinqiao Rd. 451, Shanghai, 200136, China}

\author{Tuo Ji}
\affiliation{Polar Research Institute of China, Jinqiao Rd. 451, Shanghai, 200136, China}
\affiliation{Key Laboratory for Research in Galaxies and Cosmology, The University of Science and Technology of China, Chinese Academy of Sciences,
Hefei, Anhui, 230026, China}

\author{Shaohua Zhang}
\affiliation{Polar Research Institute of China, Jinqiao Rd. 451, Shanghai, 200136, China}
\affiliation{Key Laboratory for Research in Galaxies and Cosmology, The University of Science and Technology of China, Chinese Academy of Sciences,
Hefei, Anhui, 230026, China}

\author{Xiheng Shi}
\affiliation{Polar Research Institute of China, Jinqiao Rd. 451, Shanghai, 200136, China}
\affiliation{Key Laboratory for Research in Galaxies and Cosmology, The University of Science and Technology of China, Chinese Academy of Sciences,
Hefei, Anhui, 230026, China}

\author{Ji Wang}
\affiliation{Department of Astronomy, California Institute of Technology, MC 249-17, 1200 E. California Boulevard, Pasadena, CA 91106, USA}

\author{Ji-Lin Zhou}
\affiliation{School of Astronomy and Space Science, Nanjing University, Nanjing, Jiangsu, 210093, China}
\affiliation{Key Laboratory of Modern Astronomy and Astrophysics (Nanjing University), Ministry of Education, Nanjing, Jiangsu, 210093, China}

\author{Hongyan Zhou}
\affiliation{Polar Research Institute of China, Jinqiao Rd. 451, Shanghai, 200136, China}
\affiliation{Department of Astronomy, The University of Science and Technology of China, Hefei, Anhui, 230026, China}
\affiliation{Key Laboratory for Research in Galaxies and Cosmology, The University of Science and Technology of China, Chinese Academy of Sciences,
Hefei, Anhui, 230026, China}

\correspondingauthor{Peng Jiang, Hui-Gen Liu}
\email{jpaty@mail.ustc.edu.cn, huigen@nju.edu.cn}

\begin{abstract}
Proxima~Centauri is known as the closest star to the Sun. Recently, radial velocity(RV) observations
revealed the existence of an Earth-mass planet around it. With an orbital period of $\sim$11 days. Proxima Centauri b is probably in the habitable zone of its host star. We undertook a photometric monitoring campaign to search for its transit, using the Bright Star Survey Telescope at the Zhongshan Station in Antarctica.
A transit-like signal appearing on 2016 September 8 has been tentatively identified. Its midtime,
$T_{C}=2,457,640.1990\pm0.0017$~HJD, is consistent with the predicted ephemeris based on the RV orbit in a 1$\sigma$
confidence interval. Time-correlated noise is pronounced in the light curve of Proxima~Centauri, affecting the
detection of transits. We develop a technique, in a Gaussian process framework, to gauge the statistical significance
of a potential transit detection. The tentative transit signal reported here has a confidence level of $2.5\sigma$.
Further detection of its periodic signals is necessary to confirm the planetary transit of Proxima~Centauri~b.
We plan to monitor Proxima~Centauri in next Polar night at Dome~A in Antarctica, taking the advantage of continuous darkness.
\citet{Kipping17} reported two tentative transit-like signals of Proxima Centauri b, observed by the Microvariability
and Oscillation of Stars space telescope in 2014 and 2015, respectively. The midtransit time of our
detection is 138 minutes later than that predicted by their transit ephemeris. If all of the signals are real transits,
the misalignment of the epochs plausibly suggests transit timing variations of Proxima~Centauri~b induced by an outer
planet in this system.
\end{abstract}


\section{Introduction}
The detection of terrestrial exoplanets is important for studying the population, diversity, and habitability
of planets beyond the solar system. Limited by atmospheric scintillation \citep{Dravins98}, ground-based
telescopes cannot deliver a photometric precision comparable to space-base telescopes, which is required
to detect transits of terrestrial planets around solar-like stars. Red dwarfs have much smaller size
than the Sun, i.e. $R \leq 0.15 R_{\sun}$. The transits of Earth-size planets around red dwarfs, with a
typical depth of $\sim$0.5\%, are considerable for observation using ground-based telescopes
(e.g., \citealt{Char09,Gillon16,Gillon17}). Aside from their small sizes, the masses of red
dwarf planet hosts are small. Periodic radial velocity (RV) modulations of red dwarfs induced by their
terrestrial planets are observable using current spectrographic technique with a precision of $\sim$ 1 m $s^{-1}$
\citep{Bouchy01, Pepe11, Fischer16}. Moreover, red dwarfs are abundant near the Sun. The survey conducted
by the REsearch Consortium On Nearby Stars (RECONS; \citealt{Henry06}) shows that $\sim$69\% of stars
within 10~pc are red dwarfs.

Proxima Centauri, a red dwarf, is well-known as the nearest star to the Sun, with a distance
of 1.3008$\pm0.0006$pc \citep{Kervella16}. Proxima Centauri and $\alpha$ Centauri A and B are in a
hierarchical triple star system. $\alpha$ Centauri A and B constitute a binary subsystem, while Proxima
Centauri is currently close to the apastron (13.0$^{+0.3}_{-0.1}$~kAU) of its orbit ($e=0.5^{+0.08}_{-0.09}$)
around $\alpha$~Centauri AB \citep{Kervella17}. Planets around these three stars, if discovered, would
 plausibly be the closest exoplanets to the Earth.

Firstly, an Earth-mass planet $\alpha$ Centauri Bb was claimed by \citet{Dumusque12} based on the
RV measurements using the High Accuracy Radial velocity Planet Searcher (HARPS).
However, the weak planetary signal with a semi-amplitude of $\sim0.5$~m $s^{-1}$ is most likely spurious,
arising from the time sampling of the original data and intrinsic stellar activity \citep{Rajpaul15, Rajpaul16}.
Analyzing the light curve observed using the Hubble Space Telescope (HST) over 40 hr, \citet{Demory15} detected
a single transit-like event but it is unlikely associated with $\alpha$~Centauri~Bb. Recently, an
Earth-mass planet around Proxima Centauri was discovered, using two high-precision RV instruments
\citep{Guillem16}, HARPS and the Ultraviolet and Visual Echelle Spectrograph (UVES) jointly. With an orbital
period of $11.186^{+0.001}_{-0.002}$ day, the surface of Proxima~Centauri~b is temperate and might be habitable.

The geometric probability of transit for Proxima~Centauri~b is just about 1.5\%. Nevertheless, efforts
to detect its planetary transit are valuable, as such a discovery would be very important. The
planetary mass and radius can be accurately measured, enabling us to study its structure and internal
compositions \citep{Fortney07, Dressing15}. Furthermore, its atmosphere can be characterized using transmission
spectroscopy technique. The habitability of a planet is very sensitive to the properties and
compositions of its atmosphere \citep{Ribas16, Turbet16, Meadows16, KL16}. The transit depth of
Proxima~Centauri~b would be $\sim$ 5~mmag, assuming it has a similar density to the Earth \citep{Guillem16}.
The required photometric precision can be easily achieved with small aperture telescopes for bright stars, and
thus searching for the planetary transit of Proxima Centauri b is affordable.

The first attempt was made by \citet{Kipping17} using the archival data of the Microvariability and
Oscillation of Stars space Telescope ($MOST$) observed in 2014 (for 12.5 days) and in 2015 (for 31 days).
Proxima Centauri was observed for a fraction ($<$ 50\%) of each
$MOST$ 101 minute orbit when it was observable, with an average sampling rate of about 1 minute. A candidate transit signal with two events were detected marginally.
However, the signal cannot be recovered in the light curve observed by the HATSouth telescope network \citep{Bakos13}.
Proxima~Centauri is a moderately active star and has a rotation period of $\sim$83 days
\citep{Christian04,Kiraga07,Davenport16}. Time-correlated structure and long-term trends are pronounced
in its light curve \citep{Kipping17}. Considering the expected transit duration of Proxima Centauri b is
about 1 hour, the $MOST$ light curve with discrete sampling windows in a width of $<$ 50 minutes but
separated by $>$ 50 minutes, is not optimal for disentangling the potential transit signals from
time-correlated noise. High-cadence consecutive photometric monitoring are valuable for Proxima~Centauri.

We started the photometric monitoring of Proxima~Centauri in Antarctica right after the announcement of the
discovery of Proxima~Centauri~b \citep{Guillem16}. In August and September, Proxima Centauri is not observable
at almost all observatories around the world, except for the sites in Antarctica. Our observations were
carried out using the Bright Star Survey Telescope (BSST; \citealt{Tian16}) deployed at the Chinese Antarctic
Zhongshan station (south 69$^{\circ}22^{'}23^{''}$, east $76^{\circ}22^{'}17^{''}$). In this paper, the BSST
telescope is briefly introduced in \S2. Observation and data reduction are presented in \S3.
We search for transit signals in the BSST light curves in \S4. A discussion and brief summary are presented
in \S5 and \S6, respectively. 

\section{The Bright Star Survey Telescope in Antarctica}
Dome A, on the top of the ice cap in Antarctica, is one of the most promising astronomical sites on the Earth
(\citealt{Yang09}; \citealt{Bonner10}; \citealt{Shi16}). The Chinese Antarctic Kunlun station
(south 80$^{\circ}25^{'}01^{''}$, east $77^{\circ}06^{'}58^{''}$), which is 7.3 km away from Dome A in the
southwest, is currently hosting two wide-field telescopes with aperture of 500~mm (the Antarctic Survey
Telescopes; \citealt{Yuan08}) and will be developed to support more astronomical facilities in the near future.
The survey of transiting extrasolar planets, taking advantage of the continuous darkness and large clear-sky
fraction ($>$90\%; \citealt{Zou10, Law13, Wang14}) in the winter months, is one of the main scientific
goals at Kunlun station.

BSST (\citealt{Tian16}) is optimally designed for searching planetary
transits of bright stars and will join the ongoing survey at Kunlun station. It has an aperture size of
300~mm and is equipped with a large frame $4K\times4K$ CCD camera to receive starlight from a
$3.^{\circ}4 \times 3.^{\circ}4$ field of view (\citealt{Li15}). An autonomous observation and control
system developed on the basis of EPICS\footnote{Experimental Physics and Industrial Control System,
http://www.aps.anl.gov/epics/} and RTS2 (\citealt{Kub06,Kub08}) enables BSST to run robotically \citep{Zhang16}.
Low-temperature-resistant testing in the laboratory demonstrated that the telescope can work functionally in
ambient temperatures down to $-70^{\circ}$C \citep{Tian16}. In 2016 March, BSST was installed
at the Chinese Antarctic Zhongshan station, on the Larsemann Hills in Prydz Bay, for testing operation.
Its first observation season in Antarctica ceased at the end of 2016 September. We plan to move BSST
to Kunlun Station in two years.

\section{Observation and Data Reduction}
We have monitored the photometry of Proxima Centauri over 10 nights between 2016 August 29 and September 21.
The star was observed in its white light, without any filter. The typical airmass toward Proxima Centauri
was 1.4 allthrough the observation campaign. Five exposures of 3 seconds are combined as a photometric
point, and a typical cadence of data points is 220 seconds.

Raw images are first corrected for bias and flat-field using the standard procedures. We manually select 62 nearby
stars with similar brightness to Proxima Centauri for aperture photometry. Since the observed star
field is quite dense, the background for each star can hardly be determined in a simple way of using
concentric circles. We then manually select the nearest region around each star without significant starlight to estimate its background level. Aperture photometry of Proxima Centauri and the 62 preliminary
reference stars is extracted for each image. The five consecutive data points of each star are then
averaged numerically to produce the raw light curves.

%tables:
\begin{table}[htb!]
\begin{center}
%\tablewidth{0pt}
\caption{Photometry of Proxima~Centauri Observed by BSST.}
\begin{tabular}{cccc}
\hline
HJD$-$2,457,545 & Flux & Flux & Uncertainty \\
(day) & \footnote{Normalized differential flux of Proxima~Centauri} &
\footnote{Detrended flux by fitting a cubic polynomial to data observed in individual nights} &  \\
\hline
   85.1684 & 0.9956 & 0.9941 & 0.0049 \\
   85.1709 & 0.9915 & 0.9889 & 0.0049 \\
   85.1739 & 0.9969 & 0.9962 & 0.0049 \\
   85.1766 & 0.9941 & 0.9926 & 0.0049 \\
   85.1792 & 1.0053 & 1.0075 & 0.0049 \\
   85.1817 & 1.0069 & 1.0098 & 0.0049 \\
   85.1843 & 1.0070 & 1.0101 & 0.0049 \\
   85.1871 & 1.0010 & 1.0023 & 0.0049 \\
   85.1911 & 1.0081 & 1.0118 & 0.0049 \\
   85.1940 & 0.9979 & 0.9986 & 0.0049 \\
   85.1968 & 1.0015 & 1.0034 & 0.0049 \\
   85.1995 & 1.0013 & 1.0032 & 0.0049 \\
   85.2022 & 1.0008 & 1.0027 & 0.0049 \\
   85.2053 & 1.0022 & 1.0046 & 0.0049 \\
   85.2089 & 0.9954 & 0.9958 & 0.0049 \\
   ...   &  ...   &   ...  &  ... \\
\hline
\end{tabular}
\end{center}
\label{data_BSST}
\tablecomments{Time has been converted to HJD (Heliocentric Julian Day) from UTC
(Universal Time Coordinated) time. This table is available in its entirety
in machine-readable form in the online journal. A portion is shown here for guidance
regarding its form and content.}
\end{table}

We reject unstable stars among the 62 preliminary reference stars by comparing their raw light curves
mutually and iteratively. Finally, the six nearest stable stars around Proxima Centauri are selected
as reference stars for differential photometry. We further identify the photometric points suffering
from significant extinction due to clouds in the raw light curves. These points are then clipped to
produce the clean light curves. 

The differential light curve of Proxima Centauri is derived by comparing its clean light curve with
the average clean light curve of the six reference stars. Among the six stable stars, we select a star,
which has almost the same brightness as Proxima Centauri, to estimate the photometric precision independently.
The scatter of its differential light curve, which is measured by comparing the clean light
curve of the selected reference star with the average clean light curve of the other five reference stars,
is reported as the photometric precision for Proxima Centauri. Precision is estimated night by night.
Overall, the average precision is 4.6 mmag, which is significantly larger than the theoretical
photon noise of $\sim$ 1 mmag. We estimate the amplitude of the atmospheric scintillation \citep{Dravins98}
as $$\sigma = 0.09{\frac{X^{3/2}}{D^{2/3}\sqrt{2t}}\exp{\left(-{\frac{h}{8}}\right)}}, $$
where $X=1.4$ is the airmass, $D=30$ the diameter of the telescope in cm, $t=15$ the exposure time in s, and
$h=0$ the altitude of observatory in km, yielding $\sigma\approx3$~mmag. Thus, the photometric precision
is mainly limited by the atmospheric scintillation and possible systematic errors.

\begin{figure}[htb!]
\vspace{0cm}\hspace{0cm}
\centering
\includegraphics[scale=0.7]{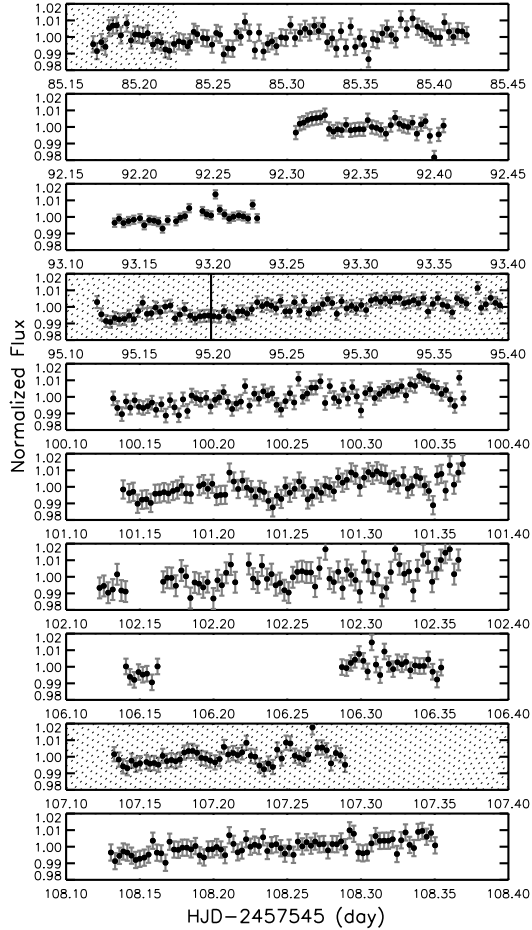}
\vspace{0cm}
\caption{Normalized differential light curve of Proxima~Centauri observed using BSST at the Antarctic
Zhongshan station during 10 nights from 2016 August 29 to September 21. Time has been converted to HJD from UTC. 
The average photometry precision is 4.6~mmag. The shaded areas are time within 1$\sigma$ predicted
transit windows ($T_{\rm IC}=2,457,629.56\pm0.67$, $T_{\rm IC}=2,457,640.74\pm0.67$ HJD, and
$T_{\rm IC}=2,457,651.93\pm0.67$). The transit ephemeris is derived by \citet{Kipping17} using the radial
velocity solution of \citet{Guillem16}. The vertical solid line indicates the midtransit time of
the candidate event identified in \S4. \label{fig1}}
\end{figure}

The normalized light curve of Proxima~Centauri is tabulated in Table \ref{data_BSST} and illustrated in
Figure \ref{fig1}. Time-correlated noise is pronounced in the BSST light curve. In order to enhance the
visibility of the potential transit features, we detrend the light curve by fitting a cubic polynomial to the
data observed in individual nights. That procedure removes the long-term variations, but can preserve
the features with short timescales as well as potential transit signals. The detrended light curve of
Proxima~Centauri is tabulated in Table \ref{data_BSST} as well. Note that the detrended light curve is
only used for visual inspection of transit features, but not for statistical analysis since time-correlated
noise has been reduced artificially.

\section{Searching for Transit Signal in BSST data}
\subsection{A Candidate Transit Event}
If Proxima~Centauri~b transits, the transit minimum would appear at the time of inferior conjunction. Using the
RV solution of \citet{Guillem16}, \citet{Kipping17} derived the time of inferior conjunction
$T_{\rm IC}=2,456,678.78\pm0.56$~HJD at an original epoch and the orbital period $P=11.1856\pm0.0013$~days.
We adopt the transit ephemeris in \citet{Kipping17} to predict the time of transit window in the BSST light curve.
The first observing night is partially covered by the 1$\sigma$ window of interest ($T_{\rm IC}=2,457,629.56\pm0.67$~HJD
for the 85th epoch). The fourth and ninth observing nights are fully covered ($T_{\rm IC}=2,457,640.74\pm0.67$~HJD
for the 86th epoch, $T_{\rm IC}=2,457,651.93\pm0.67$~HJD for the 87th epoch). The windows of interest
for the transit are indicated by the shaded regions in Figure \ref{fig1}.

Besides the transit ephemeris, the transit duration and depth are requisite parameters to characterize the
planetary transit signals in light curves. Proxima~Centauri has a stellar mass of $M_{\ast}=0.120\pm0.015 M_{\sun}$
and a stellar radius of $R_{\ast}=0.141\pm0.021 R_{\sun}$ \citep{Delfosse00, Boyajian12}. The minimum mass of
Proxima~Centauri~b is $M_{p}\sin i = 1.27^{+0.19}_{-0.17} M_{\earth}$~\citep{Guillem16}. Assuming a circular
planetary orbit, the transit duration can be expressed as
$$T_{d}=\frac{P}{\pi}\sin^{-1}\left(\frac{\sqrt{\left(R_{\ast}+R_{p}\right)^2-\left(bR_{\ast}\right)^2}}{a}\right)$$,
where $b$ is the impact parameter, $R_{p}$ is the planetary radius, and
$a\approx\sqrt[^3\!]{\frac{GM_{\ast}T^2}{4\pi^2}}=0.048$~AU is the semi-major axis.
If Proxima~Centauri~b transits, its true mass would be almost identical to the minimum mass
$M_{p}\approx1.27M_{\oplus}$. Assuming that the planet has a density similar to the Earth, the planetary radius
can therefore be estimated as $R_{p}\approx1.08R_{\oplus}$. Then, the planetary transit duration is about 75
minutes when $b=0$, and is about 66 minutes when $b=0.5$. The planetary transit depth is
$D=\left({R_p}/{R_{\ast}}\right)^2=0.48\%$.

\begin{figure}[htb!]
\vspace{0cm}\hspace{0cm}
\centering
\includegraphics[scale=0.8]{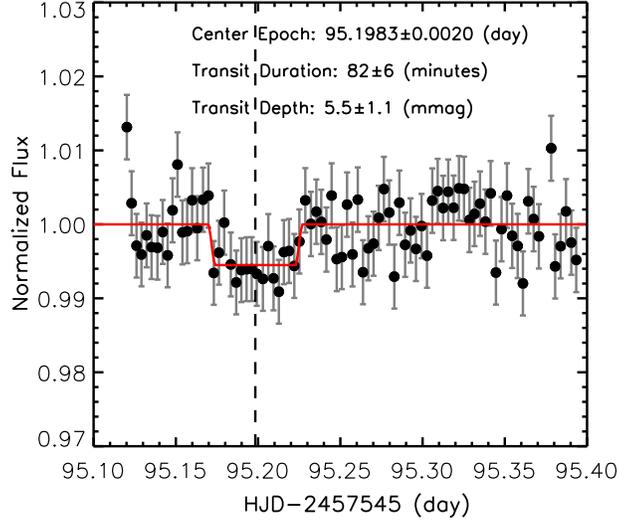}
\vspace{0cm}
\caption{Candidate transit event in the detrended light curve. It is in the 1$\sigma$ window of interest
of $T_{\rm IC}=2,457,640.74\pm0.67$~HJD for the 86th epoch. The best-fitted model, with
$T_{\rm C}=2,457,640.1983\pm0.0020$~HJD, $T_{d}=82\pm6$~minutes, and $D=5.5\pm1.1$~mmag, is overplotted
(the red solid curve). \label{fig2}}
\end{figure}

We inspect the detrended BSST light curve of Proxima~Centauri to look for features similar to
the predicted transit signal. A candidate transit event at the epoch of $\sim$ 2,457,640.2~HJD is identified. It
is in the 1$\sigma$ window of interest of $T_{\rm IC}=2,457,640.74\pm0.67$~HJD for the 86th epoch (see Figure
\ref{fig1}). We fit the candidate event using a simple transit model assuming the star is a uniform source
\citep{MA02}. There are three free parameters in the model, i.e. the time of transit minimum $T_{\rm C}$,
transit duration $T_{d}$ and transit depth $D$. The fit is carried out with the Levenberg-Marquardt
least-squares algorithm
\footnote{IDL procedures developed by Craig B. Markwardt, http://cow.physics.wisc.edu/~craigm/idl/idl.html},
yielding $T_{\rm C}=2,457,640.1983\pm0.0020$~HJD, $T_{d}=82\pm6$~minutes, and $D=5.5\pm1.1$~mmag. The best-fitted
transit parameters are compatible with the estimated parameters of the transit signal induced by Proxima~Centauri~b,
considering the uncertainty of the stellar parameters and the parameters of planetary orbit. The candidate
transit event and its best-fitted model are presented in Figure \ref{fig2}.

\subsection{Time-correlated Noise and Statistical Significance}
Time-correlated noise, induced by changing atmospheric conditions, and instrumental and stellar variability,
has a very important effect on the detection threshold of planetary transit surveys \citep{Pont06}.
Proxima~Centauri exhibits pronounced time-correlated structure in its light curve. Thus, it is necessary
to take correlated noise into account for gauging the statistical significance of the candidate
transit signal.

We estimate the time-correlated noise $\sigma_{\rm tc}$ in the BSST normalized light curve, following the
procedure introduced by \citet{Pont06}. We calculate the mean flux $F_{j}$ over a sliding interval of length
of the transit duration obtained in \S4.1. The variance of $F_{j}$ is then considered as
the time-correlated noise, yielding $\sigma_{\rm tc}=2.5$~mmag.\footnote{$\sigma_{\rm tc}$
consists of white noise, $\sigma_{w}$, and time-correlated noise. In this case, the sliding windows generally
have 16 data points. Then, the white noise can be estimated as $\sigma_{w}=\sigma_{s}/\sqrt{16}=1.15$~mmag, where
$\sigma_{s}=4.6$~mmag is the average precision of a single photometric datum. Thus, $\sigma_{\rm tc}$
is mostly contributed by time-correlated noise.} The statistical significance for the candidate transit
detection is therefore $S_{\rm tc} = D/\sigma_{\rm tc} = 5.5/2.5 = 2.2$, where $D=5.5$~mmag the transit
depth obtained in \S4.1.

\subsection{A Gaussian Process Framework}
Time-correlated noise in astronomical data usually has a power spectral density varying as
$f^{-\alpha}$~\citep{Collier01, Carter09}. In the scheme of \citet{Pont06}, the noise
$\sigma_{\rm tc}$~is estimated as the cumulative time-correlated noise over all timescales
appearing in the light curve. Thus, $\sigma_{\rm tc}$~is dominated by the fluctuations at long
timescales. However, the time-correlated noise at timescales longer than the transit duration
(i.e. a typical timescale of 1--3 hr) has little impact on the planetary transit signals.
This is because transit signals can be disentangled from the correlated noise at long timescales
effectively. Note that the long-term trend is an extreme case of time-correlated noise at long timescales.
Moreover, the noise at short timescales has small amplitudes as $\sigma_{\rm tc} \propto f^{-\alpha}$,
which hardly affects the detection threshold. Therefore, the blurring of transit signals is mainly
attributed to correlated noise at timescales comparable to the transit duration, if time-correlated
noise were the dominant noise source.

Gaussian process (GP) models define a distribution over function space, allowing each point in the time
series to have some degree of correlation with every other point. GP regression has been used to model
the systematics in a non-parametric way for planetary transit and RV observations
(e.g., \citealt{Gibson12, Rajpaul15, Kipping17}). Furthermore, GP regression is an interpolation, as well
as a fitting, technique. It can be used to reconstruct noisy, irregularly sampled data \citep{Rybicki92, Press07}.
In the following, we gauge the significance of the candidate transit event of Proxima~Centauri~b observed
in the BSST light curve in a GP framework.

The full normalized light curve is regressed with GP to model the correlated noise.\footnote{We performed GP
regression using the Python package developed by \citet{Gibson12}. The package is accessible through
https://github.com/nealegibson/GeePea} Two popular GP kernels, the Mat{\'e}rn 3/2 kernel and
Squared Exponential kernel, are adopted for regression. There are three hyperparameters,
$\theta_{\rm hyper}=(\alpha, l, \sigma)$, where $\alpha$~controls the magnitude (in units of mmag), $l$
controls the timescale of correlations (in units of minutes), and $\sigma$ represents the uncertainties of
data points (in units of mmag). Regression yields
$\theta_{\rm hyper}=(3.31^{+0.37}_{-0.32},29.86^{+8.31}_{-6.67},4.05^{+0.15}_{-0.14})$
for the Mat{\'e}rn 3/2 kernel and
$\theta_{\rm hyper}=(3.20^{+0.34}_{-0.32},23.54^{+4.24}_{-3.67},4.11^{+0.13}_{-0.13})$
for the Squared Exponential kernel.

\begin{figure}
\vspace{0cm}\hspace{0cm}
\centering
\includegraphics[scale=0.7]{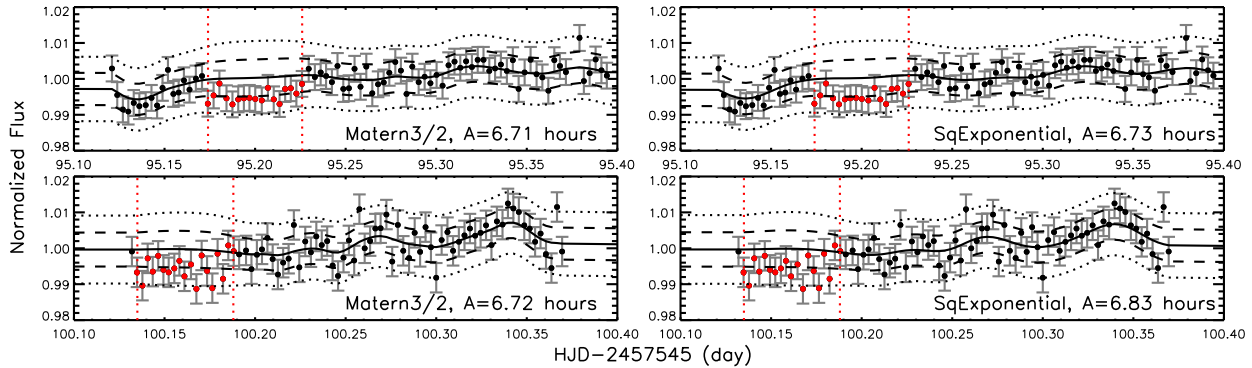}
\vspace{0cm}
\caption{Illustration of the GP framework. Two windows with a time length of 80 minutes are presented.
The left panels are the GP regressions using the Mat{\'e}rn 3/2 kernel and the right panels are for the Squared
Exponential kernel. In all panels, the blocked data points are in red, the solid line is the best
prediction of GP regression, and the dashed lines and dotted lines are the 1$\sigma$ and 2$\sigma$ confidence
regions, respectively. The integrated depths of these two windows are the largest among those of all the
80 minutes windows. The window in the top panels matches the candidate transit event of Proxima~Centauri~b,
having an integrated depth of $A_{M}=6.71$ hr for the Mat{\'e}rn 3/2 kernel, while $A_{\rm SE}=6.73$ hr for the
Squared Exponential kernel. The window in the bottom panels has $A_{M}=6.72$~hours and $A_{\rm SE}=6.83$ hr.
\label{fig3}}
\end{figure}

We then interrupt the full light curve by blocking data points in sliding windows with varying lengths.
Thus, the interrupted light curves have different midtimes of blocked windows and different window lengths
from each other. For a specific window length, over 500 interrupted light curves are produced.
We perform GP regression as an interpolation to predict the flux at the original time stamps in the blocked windows
of the light curves, while the hyperparameters are fixed to those obtained in the full light curve. The predicted
flux is considered as the best linear unbiased prediction of the intermediate flux under the priors of the GP
model fit to the light curve \citep{Press07}. We define an integrated depth $A=\int^{t_1}_{t_0}\left(F_{\rm pre} - F_{\rm obs}\right)dt$,
where $t_0$ and $t_1$ are the edges of the blocked windows, $F_{\rm pre}$ is the predicted flux, and $F_{\rm obs}$ is the observed
flux. For the interrupted light curves with the same window length, the variation of $A$ quantifies the
fluctuation induced by correlated noise at timescales comparable to the window length. Although correlated noise
at shorter timescales and the white noise contribute to the variation of $A$ as well, their amplitudes are smaller.

\begin{figure}
\vspace{0cm}\hspace{0cm}
\centering
\includegraphics[scale=0.7]{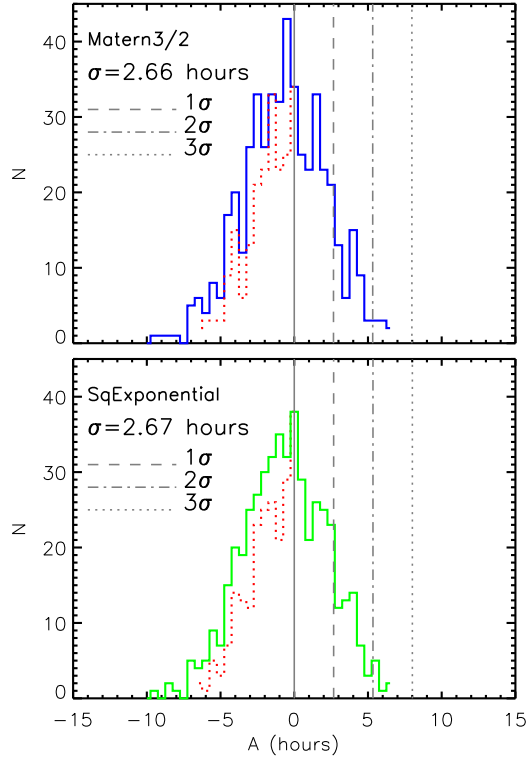}
\vspace{0cm}
\caption{Distribution of integrated depths $A$ for all of the windows with a time length of 80 minutes.
GP regressions using the Mat{\'e}rn 3/2 kernel (blue) and Squared Exponential kernel (green) give similar distributions.
The integrated depths of $A > 0$ represent the depression of the observed flux compared to the GP predicted flux, while those with
$A < 0$ represent the the excess. The distribution is asymmetric, as the frequent stellar flaring of Proxima~Centauri
enhances the probability of flux excess. The integrated depths of $A > 0$ are mirrored (red dotted curve) to construct
a nominal full distribution. Its standard deviation $\sigma$ is reported as the fluctuation purely induced by
time-correlated noise, yielding $\sigma_{M}=2.66$ hr (Mat{\'e}rn 3/2 GP models) and $\sigma_{\rm SE}=2.67$ hr
(Squared Exponential GP models) for the time length of 80 minutes.
\label{fig4}}
\end{figure}

In order to illustrate the GP framework, we present two interrupted light curves with blocked windows with length
of 80 minutes in Figure \ref{fig3}. Both GP models with the Mat{\'e}rn 3/2 kernel and Squared Exponential kernel
are presented. The blocked data points are in red. The solid line is the best prediction from GP regression, and the
dashed lines and dotted lines are the 1$\sigma$ and 2$\sigma$ confidence regions, respectively. The window
in the top panels is the one that matches the candidate transit event of Proxima~Centauri~b. It has an
integrated depth of $A_{M}=6.71$ hr for the Mat{\'e}rn 3/2 kernel, while $A_{\rm SE}=6.73$ hr for the Squared
Exponential kernel. The window in the bottom panels has $A_{M}=6.72$ hr and $A_{\rm SE}=6.83$ hr.
It is not associated with possible transits of Proxima~Centauri~b, since its midtime
is far away from the predicted transit epochs. The integrated depths of these two windows are the largest among
those of the sliding windows with the same length (80 minutes).

\begin{figure}
\vspace{0cm}\hspace{0cm}
\centering
\includegraphics[scale=0.7]{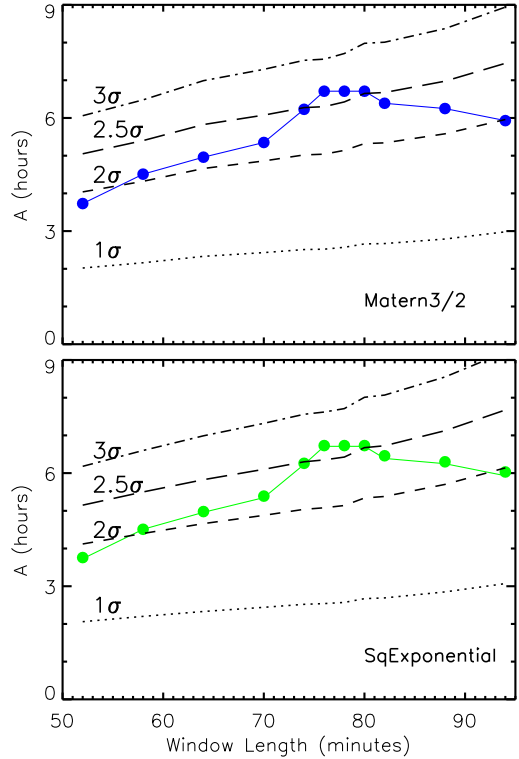}
\vspace{0cm}
\caption{Standard deviations of the integrated depths for different window lengths (dotted and dashed curves) and
the integrated depths of different-size windows centered at the midtime of the
candidate transit event (color points). The integrated depths peak at the time lengths of 76, 78, and
80 minutes, which are close to the transit duration, in our grid of GP analysis. These
three windows contains the same photometric data points, thus they are identical.
That window matches the candidate transit event in midtime and length. Its
integrated depth is over the 2.5$\sigma$ threshold in both the Mat{\'e}rn 3/2 GP
model and Squared Exponential GP model cases.
\label{fig5}}
\end{figure}

The distribution of the integrated depths for all of the windows with length of 80 minutes is presented in Figure \ref{fig4}.
GP regressions using the Mat{\'e}rn 3/2 kernel and Squared Exponential kernel give similar distributions. The integrated
depth $A > 0$~is attributed to the depression of the observed flux compared to the predicted flux, while $A < 0$ is for the
excess of observed flux comparing to predicted flux. The integrated depths distribute more widely in the range of
$A < 0$. The asymmetry is expected, as Proxima~Centauri shows frequent stellar flaring \citep{Davenport16}, enhancing
the probability of flux excess. Thus, only the integrated depths distributed in the range of $A > 0$ can be used
to gauge the fluctuation purely induced by time-correlated noise. We mirror the integrated depths of $A > 0$~to
construct a nominal full distribution and report its standard deviation as the fluctuation induced by correlated noise.
For the time length of 80 minutes, the standard deviation is $\sigma_{M}=2.66$ hr for the GP regressions using the
Mat{\'e}rn 3/2 kernel and $\sigma_{\rm SE}=2.67$ hr for the Squared Exponential kernel.

The standard deviations of the integrated depths for different window lengths are presented in Figure \ref{fig5}.
They increase with window lengths, as the correlated noise with longer timescales have larger amplitudes.
We then collect the different-size windows centered at the midtime of the candidate transit event of Proxima~Centauri~b.
Their integrated depths peak at the time lengths close to the transit duration (as seen in Figure \ref{fig5}).
In the grid of our GP analysis, the peak integrated depths appear at the window lengths of 76, 78, and 80 minutes.
These three windows contains the same photometric data points, thus they are identical.\footnote{The time resolution
is limited by the sampling rate of the light curve.}
The window matches the candidate transit event in midtime and length. It is illustrated in Figure \ref{fig3}.
Its integrated depth is $A_{M}=6.71$ hr for the Mat{\'e}rn 3/2 GP model, while $A_{\rm SE}=6.73$ hr for the Squared Exponential
GP model. In Figure \ref{fig5}, the standard deviations read $\sigma_{M}=2.52$ hr and $\sigma_{\rm SE}=2.54$ hr at
the window length of 76 minutes, $\sigma_{M}=2.57$ hr and $\sigma_{\rm SE}=2.57$ hr at the window length of 78 minutes, and
$\sigma_{M}=2.66$ hr and $\sigma_{\rm SE}=2.67$ hr at the window length of 80 minutes.
It is straightforward to define the statistic $S=A/\sigma$ as the statistical significance of the transit detections.
Conservatively, we report the lowest detection statistic $S=A_{\rm SE}/\sigma_{\rm SE}=6.73/2.67=2.52$, where the integrated
depth and standard deviation at the window length of 80 minutes in the Squared Exponential GP models are employed for
the candidate transit event in the BSST light curve. Assuming the distribution of integrated depths is
Gaussian, we expect to see two other windows with integrated depths statistically comparable to that of the tentative transit signal
among the 500 windows in the analysis. We do see one as shown in the bottom panels of Figure \ref{fig3}.
The candidate transit signal of Proxima~Centauri~b is identified by the consistency of its midtime and the predicted transit
epochs based on RV solution.

\begin{figure}
\vspace{0cm}\hspace{0cm}
\centering
\includegraphics[scale=0.7]{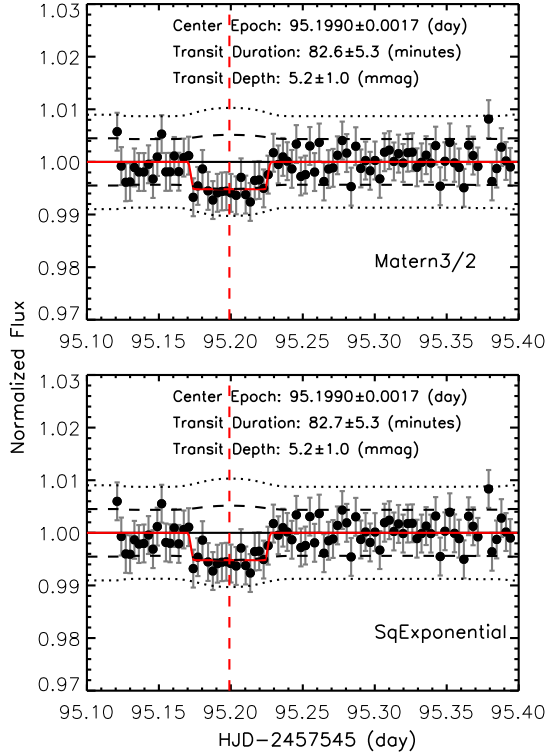}
\vspace{0cm}
\caption{Nominal fit to the candidate transit event of Proxima~Centauri~b. The observed flux is normalized
by the predicted flux from GP regression. The best-fitted transit model (red solid line) yields
$T_{\rm C}=2,457,640.1990\pm0.0017$~HJD (the vertical red dashed line), $T_{d}=82.6\pm5.3$~minutes,
and $D=5.2\pm1.0$~mmag for GP regression using the Mat{\'e}rn 3/2 kernel, while $T_{\rm C}=2,457,640.1990\pm0.0017$~HJD,
$T_{d}=82.7\pm5.3$~minutes, and $D=5.2\pm1.0$~mmag for GP regression using the Squared Exponential kernel.
The dashed lines and dotted lines are the 1$\sigma$ and 2$\sigma$ confidence regions of the GP regressions, respectively.
\label{fig6}}
\end{figure}

We further normalize the observed light curve using the GP predicted light curve and present the candidate transit
event in Figure \ref{fig6}. The best-fitted transit model yields $T_{\rm C}=2,457,640.1990\pm0.0017$~HJD,
$T_{d}=82.6\pm5.3$~minutes, and $D=5.2\pm1.0$~mmag for GP regression using the Mat{\'e}rn 3/2 kernel, while
$T_{\rm C}=2,457,640.1990\pm0.0017$~HJD, $T_{d}=82.7\pm5.3$ minutes, and $D=5.2\pm1.0$ mmag for GP regression using
the Squared Exponential kernel. Note that the fit is just nominal, as correlated noise is not considered fully. We
also attempt to fit the observed flux directly with a full model \citep{Gibson12, Kipping17}, which is a planetary
transit model joined with a GP to model the correlated noise. However, the parameters of the transit model are not
converged. The divergence is understandable, since the detection of the transit event is just tentative, with a confidence
level of $\sim2.5\sigma$.

\section{Discussion}
\subsection{Follow-up Observations}
The detection of the transit event of Proxima~Centauri~b, in this work, is tentative, at a confidence
level of $2.5\sigma$. Therefore, four such detections of transit, at least, are
required to solidify the transit of Proxima~Centauri~b at a $5\sigma$ confidence level. Consecutive high-cadence
 observations are recommended in order to disentangle reliably the transit signals from correlated noise
and frequent stellar flaring. Considering the relatively large uncertainty of the predicted
inferior conjunction \citep{Guillem16, Kipping17}, the 24 hr before and after the predicted transit epochs
are the windows of interest.

The Antarctic Survey Telescopes (ASTs) deployed at Dome~A (the Chinese Antarctic Kunlun station) in
Antarctica are the best choice for follow-up observation of Proxima~Centauri on the ground. The star is
continuously observable in polar nights, and the large clear-sky fraction ($>$90\%) at that site can guarantee
nearly uninterrupted monitoring. We plan to observe Proxima~Centauri in the next winter at Dome~A.
An observation in four sections, with a length of two days for each, is proposed.

\begin{table}[htb!]
\begin{center}
\caption{Possible Detection of Transit Timing Variations}
\begin{tabular}{ccccc}
\hline
$N_{\rm tr}$ & $T_{\rm obs}$ & $T_{\rm lin}$ & $\Delta T$ (C$-$O)\\
                  &   \multicolumn{2}{c}{HJD - 2,450,000 (day)} & minutes  \\
\hline
11 & 6,801.0594$^{+0.0053}_{-0.0042}$ & 6,801.0439 & $-22.5^{+7.6}_{-6.0}$  \\
43 & 7,159.0514$^{+0.0046}_{-0.0048}$ & 7,159.0786 & $+39.3^{+6.6}_{-6.9}$  \\
86 & 7,640.1990$^{+0.0017}_{-0.0017}$ & 7,640.1876 & $-16.8^{+2.4}_{-2.4}$  \\
\hline
\end{tabular}
\end{center}
\label{TTV}
\tablecomments{$T_{\rm obs}$~is the time of transit minimum assuming all of the candidate events in $MOST$ and BSST
light curves are real transits. $T_{\rm obs}$~is fitted linearly, yielding a period of $P=11.18858$~days.
$T_{\rm lin}$~is the best-fitted linear epoch. $\Delta T$~is the residual of the linear fit, which is considered to be TTVs.}
\end{table}

\subsection{Comparing with the Candidate Transit Events in $MOST$ Data}
Analyzing the $MOST$ light curve, \citet{Kipping17} detected two candidate transit events of
Proxima~Centauri~b with the midtransit time of $2,456,801.0594^{+0.0053}_{-0.0042}$~HJD (for the 11th epoch) and
$2,457,159.0514^{+0.0046}_{-0.0048}$~HJD (for the 43rd epoch), with an updated orbital period of
$P=11.18725^{+0.00012}_{-0.00016}$~days (see Model~2 in their Table~4). Using this ephemeris, we derive the
time of transit minimum for the 86th epoch $t_{\rm IC}=2,457,640.1032^{+0.0098}_{-0.0117}$~HJD. This is
138~minutes earlier than the candidate transit event detected in BSST data. Although all of the detections are
tentative, it is still interesting to discuss the incoherence, which may be of some help in follow-up
observations. If all of the candidate transit events of Proxima~Centauri~b are real, we interpret
the misalignment of their epochs due to transit timing variations (TTVs, \citealt{Agol05, Holman05})
induced by an outer planet in this system. We fit the three observed transit epochs linearly to derive an
optimal orbital period, yielding $P=11.18858$~days. The residuals are therefore the TTVs,
\{$-22.5^{+7.6}_{-6.0}$, $+39.3^{+6.6}_{-6.9}$, $-16.8^{+2.4}_{-2.4}$\} minutes for the \{11th, 43rd, 86th\}
epochs, as shown in Table \ref{TTV}. We adopt a half of the peak-to-valley value as the strength of the observed
TTVs, i.e. $\rm TTV_{\rm obs}\sim30$~minutes for Proxima~Centauri~b.

\begin{figure}
\vspace{0cm}\hspace{-0.3cm}
\centering
\includegraphics[scale=0.23]{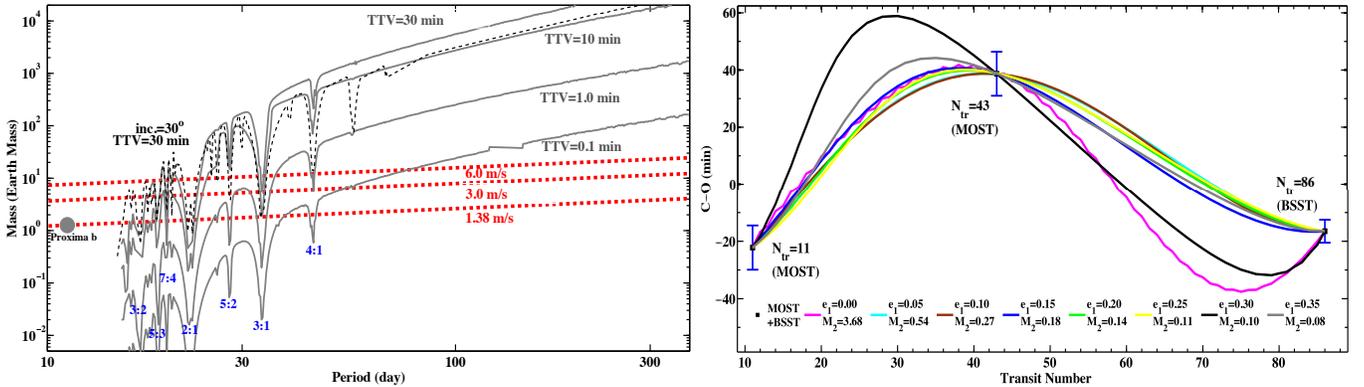}
\vspace{0cm}
\caption{Simulations of TTVs of Proxima~Centauri~b due to an outer planet. Left panel:
TTVs due to an outer planet with various masses and periods. The gray solid lines from top to bottom
represent TTV$=$~30, 10, 1, 0.1~minutes, respectively. In these simulations, we assume a coplanar orbital
configuration of the two planets, where the eccentricity of Proxima~Centauri~b is set to $e=0.1$ and
the outer planet is in a circular orbit. The black dashed line is a non-coplanar (i.e. inclination= 30$^\circ$)
model, producing TTV$=30$~minutes. The red dotted lines represent reflex stellar RV amplitudes induced by
the outer planet. Note that any planet with RV amplitude $>$3 m $s^{-1}$ can be ruled out by the RV observation of
\citet{Guillem16}. Right panel: 8 arbitrary TTV models fit to the misalignment of the transit epochs
observed by $MOST$ and BSST. In these models, the outer planet is in a coplanar near 2:1 mean motion
resonance orbit with Proxima~Centauri~b, and its period is set to 22.58509~days arbitrarily. The curves with different
colors represent best-fitted models for different designated eccentricities of Proxima~Centauri~b,
varying from $e=0$~to~$e=0.35$ in a step sizes of 0.05. The best-fitted masses of the outer planet are
listed in the legend.
\label{fig7}}
\end{figure}

Assuming a moderate eccentricity $e=0.1$ of Proxima~Centauri~b \citep{Xie16, Brown17},
we simulate the strengths of its TTVs perturbed by an outer planet with varying masses and orbital periods in
coplanar orbits. The mass and orbital period of the outer planet are constrained by the requirement that
the Doppler reflex stellar RV should be $<3$~m $s^{-1}$ to keep the planet undetectable in the high-precision
observation of \citet{Guillem16}. The simulation results are presented in the left panel of Figure \ref{fig7}.
Strengths of TTVs are significantly enhanced when the planets are nearly in mean motion resonant (MMR) orbits
\citep{Xie13,Xie14}.\footnote{Note that the planets of TRAPPIST-1 in near MMR orbits exhibit substantial TTVs \citep{Gillon17}.}
An Earth-mass planet in orbit near 2:1 or 3:2 MMR with Proxima~Centauri~b is able to produce TTVs $\ga$ 30 minutes
while keeping its reflex stellar RV~$<3$~m $s^{-1}$ . Since there are only three tentative transits in this work, further
analysis of the mass and orbital elements of the unseen planet, by fitting the observed TTV pattern, is nearly impractical.
The situation is worse in this case, as the eccentricity of Proxima~Centauri~b is quite uncertain
($e<0.35$; \citealt{Guillem16}) either. However, it is worth constructing some arbitrary TTV models to fit to the
misalignment of the transit epochs to illustrate the feasibility of the TTV scenario. The outer planets in the arbitrary models,
having an orbital period of 22.58509~days, are in coplanar near 2:1 MMR orbits with Proxima~Centauri~b. Each model has a
different designated eccentricity of Proxima~Centauri~b. In the fitting procedure, we constrain the eccentricity of
the outer planet to be small ($<0.05$), while its orbital phase difference with Proxima~Centauri~b is a free parameter.
The best-fitted models for different designated eccentricities are then presented in the right panel of Figure \ref{fig7}.
Generally, the best-fitted mass of the outer planet decreases while the eccentricity of Proxima~Centauri~b increases.
The relationship could be explained by the TTV strength increasing with the eccentricity of Proxima~Centauri~b
when the mass and orbit of the outer planet are set.

\begin{figure}
\vspace{0cm}\hspace{0cm}
\centering
\includegraphics[scale=0.26]{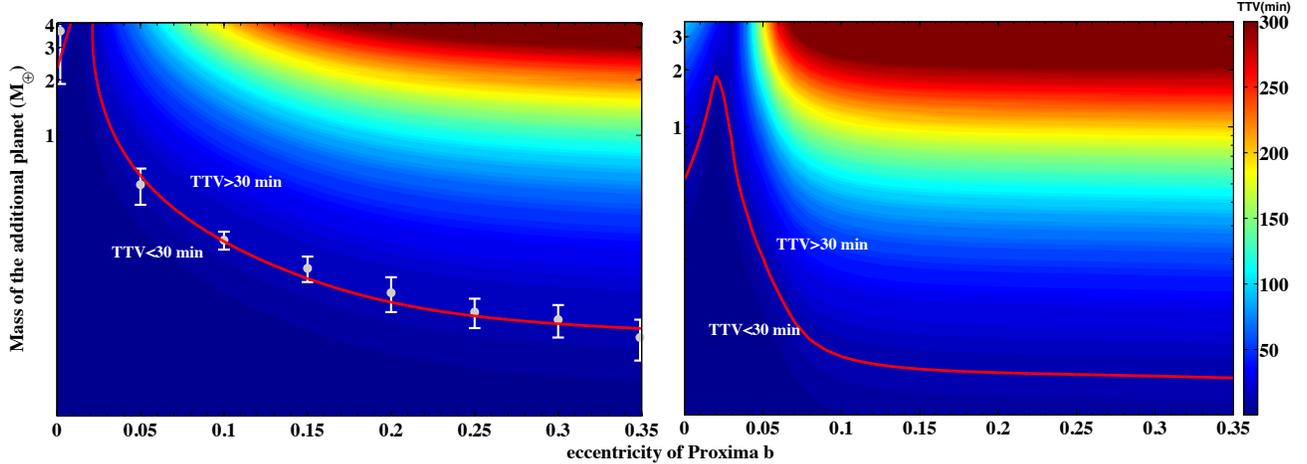}
\vspace{0cm}
\caption{Strengths of simulated TTVs of Proxima Centauri b varying with its
eccentricity and the mass of the unseen outer planet, in coplanar near 2:1 and 3:2 MMR orbital configurations.
The ranges of the eccentricity and the mass are contained by the RV observation of \citet{Guillem16}. The upper
limit of the unseen planet is 4.08~$M_{\earth}$~and 3.60~$M_{\earth}$~for near 2:1 and 3:2 MMR orbital configurations,
respectively. In the left panel, the outer planet has an arbitrary orbital period of 22.58509~days, which is in near
2:1 MMR with Proxima~Centauri~b. In the simulations, the eccentricity of the outer planet is set to zero.
The red solid curve represents the TTVs in strength of 30~minutes. The eight filled circles with error bars
represent the arbitrary TTV models in the right panel of Figure \ref{fig7}. The right panel is for the coplanar
near 3:2 MMR orbital configurations, where the orbital period of the outer planet is set to 16.89480~days arbitrarily.
\label{fig8}}
\end{figure}

Furthermore, we look over the TTV strength varying with the eccentricity of Proxima~Centauri~b,
in near 2:1 and 3:2 MMR orbital configurations. Firstly, we simulate TTVs of Proxima~Centauri~b, where the outer
planet is in a coplanar near 2:1 MMR orbit. We still use the arbitrary orbital period of 22.58509~days for
the outer planet, and set its eccentricity to zero. The mass of the planet in the outer orbit
should be less than 4.08 $M_{\earth}$ as the RV constraint \citep{Guillem16}. We present the strengths of simulated
TTVs in the left panel of Figure \ref{fig8}. For any specific TTV strength, the required mass of the outer planet decreases
with the increasing eccentricity of Proxima~Centauri~b in the range of $e\ga0.02$, but a reverse trend is
observed in the range of $e\la0.02$. The turning point can be explained by the two planets being in exact MMR when the
eccentricity of Proxima~Centauri~b is extremely small, and thus the TTV strengths in the range of $e\la0.02$ are mainly governed by
resonant perturbations. In the range of $e\ga0.02$, the two planets are not in exact MMR, but in near MMR. Therefore, different mechanisms
work in the two eccentricity regions. The turning point is more visible in the simulations for the near 3:2 MMR orbital configurations,
which are presented in the right panel of Figure \ref{fig8}. There, the unseen planet has an arbitrary orbital period of 16.89480~days
and an upper limit mass of 3.60~$M_{\earth}$.
%As the planets are closer in the near 3:2 MMR orbits, their interaction
%are stronger, producing larger TTVs than those in the near 2:1 MMR orbits while the eccentricity of Proxima~Centauri~b and
%the mass of the outer planet are specified.

Our brief discussion on TTVs just aims to demonstrate that the misalignment of the three tentative
transit events can be reasonably interpreted as TTVs of Proxima~Centauri~b induced by an outer unseen planet. The determination of
the mass and orbit of that planet is impractical in this case and therefore beyond the scope of this work.

\section{Summary}
Proxima~Centauri is the nearest star from the Sun, with a distance of about 1.3~pc. An Earth-mass planet
with an orbital period of $\sim$11 days around it has been revealed by high precision RV observations.
It is interesting that the surface of Proxima~Centauri~b is temperate and might be habitable. The geometric
probability of transit is about 1.5\%. If the planet transits, its radius, atmospheric properties, and
habitability can be well studied. We collected high-cadence light curves of Proxima~Centauri in 2016 August and September
using the BSST at the Zhongshan Station in Antarctica. We detected a tentative transit event at the
epoch of $\sim2,457,640.2$~HJD, which is compatible with the ephemeris of the RV orbit. Time-correlated noise
is pronounced in the light curve of Proxima~Centauri, while correlated noise at timescales comparable to
transit duration affect the detection of transit events. We develop a technique, in a Gaussian process framework,
to gauge the statistical significance of the candidate transit event in this work, yielding a confidence
level of $2.5\sigma$. Consecutive high-cadence observations are necessary to confirm the planetary transit
of Proxima~Centauri~b. Infrared photometry is preferred, as the stellar noise would be much weaker in that
band. Considering the relatively large uncertainty of the predicted inferior conjunction and
possible TTVs, we recommend extending the observing windows by one day before and after the predicted epochs.
We plan to perform follow-up observation of Proxima~Centauri in the next polar night at Dome~A in Antarctica,
where Proxima~Centauri is observable all day.

\acknowledgments
The authors appreciate the enlightening suggestions from the anonymous referee,
which helped to greatly improve the quality of this paper.
This work is supported by the Astronomical Project for the Chinese Antarctic
Inland Station, the SOC Program (CHINARE2012-02-03, CHINARE2013-02-03,
CHINARE2014-02-03, CHINARE2015-02-03, and CHINARE2016-02-03), and the National
Basic Research Program of China (2013CB834905, 2013CB834900, and 2015CB857005).
Hui-Gen Liu is supported by the National Natural Science Foundation of China (11503009, 11333002).
Peng Jiang is supported by the National Natural Science Foundation of China (11233002, U1431229).
Xingxing Huang acknowledges support from the China Postdoctoral Science Foundation (2015M582000).
Qi-Guo Tian is supported by the National Natural Science Foundation of China (11503023), the Natural
Science Foundation of Shanghai (14ZR1444100), and the Polar Science Innovation Fund for Young Scientists
of Polar Research Institute of China (CX20130201).
Hongyan Zhou is supported by the National Natural Science Foundation of China (11473025, 11421303, 11033007).
The authors express sincere appreciation to Mr. Yongxiang Tang (the leader) and all team members
at the Chinese Zhongshan Station, who made the operation of BSST in
Antarctica possible in the winter of 2016.

\end{document}